\documentclass[12pt,twoside]{article}
\usepackage{fleqn,espcrc1,epsfig}

\usepackage{graphicx}
\usepackage[figuresright]{rotating}

\newcommand{\AmS}{{\protect\the\textfont2
  A\kern-.1667em\lower.5ex\hbox{M}\kern-.125emS}}

\title{Nucleosynthesis Calculations for the Ejecta of Neutron Star 
Coalescences}

\author{S. Rosswog\address{Zentrum f\"ur Paralleles Rechnen/Angewandte
 Informatik (ZPR/ZAIK), Universit\"at zu K\"oln, Germany}
        and 
	C. Freiburghaus
	and
	F.-K. Thielemann\address{Departement f\"ur Physik und Astronomie,
Universit\"at Basel}}
       
\begin{document}

\maketitle
\begin{abstract}
We present the results of fully dynamical r-process network calculations
for the ejecta of neutron star mergers (NSMs).
The late stages of the inspiral and the final violent coalescence of a neutron
star binary have been calculated in detail using a 3D hydrodynamics code
(Newtonian gravity plus backreaction forces emerging from the emission 
of gravitational waves) and a realistic nuclear equation of state. The found
trajectories for the ejecta serve as input for dynamical r-process 
calculations where all relevant nuclear reactions (including beta-decays
depositing nuclear energy in the expanding material) are followed.\\
We find that all the ejected material undergoes r-process. For an initial 
$Y_e$ close to 0.1 the abundance distributions reproduce very accurately
the solar r-process pattern for nuclei with A $>$ 130. For lighter nuclei
strongly underabundant (as compared to solar) distributions are encountered.
We show that this behaviour is consistent with the latest observations 
of very old, metal-poor stars, despite 
 simplistic arguments that have recently been raised against the possibility
 of NSM as possible sources of Galactic r-process material.
\end{abstract}
\vspace*{-0.2cm}
\section{Hydrodynamic Calculations of Neutron Star Coalescences}
\begin{figure}
\hspace*{4cm}\epsfig{file=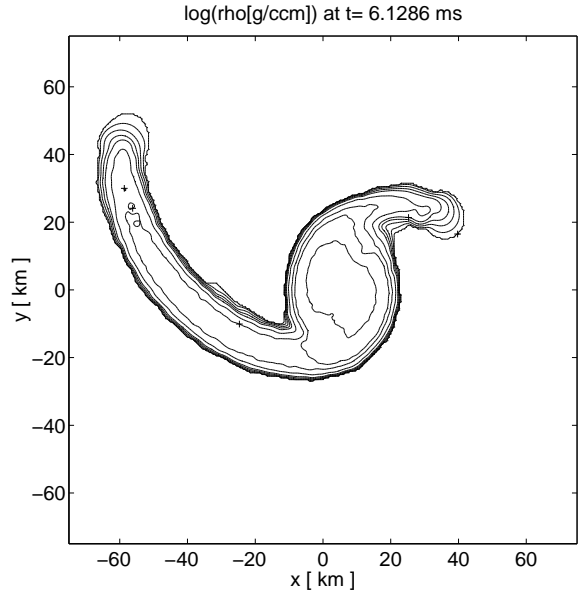,width=7.9cm,angle=90}
\caption{Density contour snap shot in the orbital plane. The  
binary system contains neutron stars of 1.3 and 1.4 M$_{\odot}$
and is initially corotating. The ejected material is located in the tip 
of the long spiral arm.}
\end{figure}
The possible impact of merging neutron stars for r-process nucleosynthesis
has been realized many years ago \cite{lattimer74,lattimer76,symbalisty82,eichler89}.
To quantify this issue we have
performed detailed hydrodynamic studies of the last inspiral stages and the 
final coalescence of neutron star binary systems. We used the smoothed particle
hydrodynamics method to solve the hydrodynamic equations coupled with
the nuclear equation of state of Lattimer and Swesty (\cite{lattimer91};
LS-EOS). To resolve shocks accurately a largely improved, hybrid artificial 
viscosity scheme \cite{rosswog2000} that profits from two recently 
suggested modifications of the 'standard scheme'
\cite{balsara95,morris97} has been applied. Gravitational forces
are evaluated efficiently using a binary tree \cite{benz90}, the backreaction
forces resulting from the emission of gravitational waves are accounted for
in the quadrupole approximation of point masses \cite{rosswog99,rosswog2000}.
For further details and test calculations we refer to 
\cite{rosswog99,rosswog2000}.\\
To explore the possible range of outcomes depending on the physical
parameters of neutron star binary systems a variety of different initial
configurations has been examinated. For example, binary systems -symmetric
and asymmetric ones- with six different initial spin combinations and 
three different neutron star masses (1.3, 1.4 and 1.6 M$_\odot$) have been 
investigated.\\
In all of these cases between $\sim 4\cdot10^{-3}$ and 
$\sim 4\cdot10^{-2}$ M$_\odot$ were ejected into space. Folded with the
estimates for the event rate of $\sim 10^{-5}$ per year and galaxy
\cite{narayan91,vandenheuvel96,fryer99}, this amount could contribute 
substantially to the enrichment of the Galaxy with heavy elements, 
provided that the ejecta contain large amounts of r-process nuclei.
\section{Fully dynamical r-process Network Calculations}

To clarify issue of the abundances within these ejecta, 
we have taken the history of a blob of ejected material located in a 
rapidly expanding spiral arm from our best resolved calculation 
(a corotating binary system) as a starting point for the  
network calculations. For two reasons $Y_e$ is treated as a free parameter
in these calculations. First, with the current resolution of our 
3D-calculations we cannot follow closely enough the steep density gradients 
in the thin, but rather proton rich neutron star surface, that determine
the initial $Y_e$ via the beta-equilibrium condition $\hat{\mu}= \mu_e(Y_e)$.
Second, $Y_e$ evolves during the coalescence process by means of weak 
interactions. Thus all the relevant weak interaction processes and neutrino
transport should be included, which is currently not the case. 
We started our nucleosynthesis calculations at a point where the density
dropped below the neutron drip density. Above this density all beta-decays 
are Pauli-blocked and thus the material cools adiabatically during the 
rapid expansion.
Along the expansion trajectory of the ejected matter blob all relevant nuclear
reactions including beta-decays have been followed \cite{freiburghaus99}. 
The temperature history within the ejecta is determined by the interplay
of cooling due to the rapid expansion and the heat input from the decaying
nuclei. To ensure that results are not influenced by possibly artificially
high numerical temperatures, we also performed test calculations starting
exclusively with neutrons and protons at an artificially low temperature
($T= 10^{-2}$ MeV). In this case the almost immediate recombination of 
the nucleons into $\alpha$-particles heated up the material to r-process-like 
temperatures on extremely short time scales \cite{freiburghaus99}.\\
In all investigated cases r-process took place, and the initial $Y_e$ turned
out to be the decisive parameter that determined the resulting abundance 
distribution. For $Y_e$-values too large or too small the abundance peaks were
shifted towards those of the s-process, for $Y_e$ close to 0.1 the solar 
abundances above A $>$ 130 are very closely reproduced. 
Isotopes with A $<$ 130 very largely underabundant compared to the solar 
values.
\begin{figure}
\hspace*{2cm}\epsfig{file=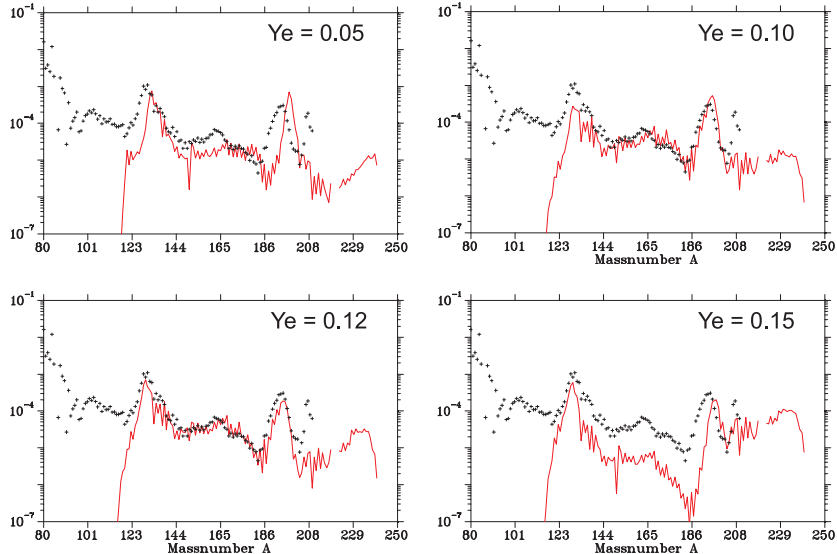,width=11cm,angle=0}
\caption{Calculated abundance patterns for different initial $Y_e$-values.}
\end{figure}
\section{Discussion}

Despite recent progress in the questions related to the nucleosynthetic impact
of neutron star mergers uncertainties remain:

\begin{itemize}
\item \underline{EOS}: While matter properties at subnuclear densities
seem to be reasonably well understood, the high density regime, 
where exotic nuclear states are expected to appear, remains rather uncertain. 
We found that the 
stiffness of the neutron star EOS has a dramatic effect on the amount of 
ejected material. For example, a corotating binary system following the 
rather stiff LS-EOS
releases $\sim 4 \cdot 10^{-2}$ M$_\odot$ in the tips of the emerging spiral
arms, while for a similar system governed by a $\Gamma= 2.0$-polytrope does not
show any resolvable mass loss \cite{rosswog99}. Test runs with polytropes
of density dependent stiffness revealed that it is unfortunately the poorly
known high density regime that determines the amount of ejected material.
Therefore, if the nuclear matter at high density should be softer (stiffer) 
than the LS-EOS less (more) material should become unbound.
Most recent EOSs \cite{akmal98,shen98} are rather stiff and therefore
support the results found with the LS-EOS.

\item \underline{Gravity}: A neutron star binary is unquestionably a highly
relativistic system. Thus, using (basically) Newtonian gravity can only
be a first approach for the exploration of  the system dynamics. Recently, 
progress has been made using approximations to full GR (e.g.
\cite{ayal99,faber99,shibata2000}).
One might suspect the stronger gravity forces of GR to reduce the
amount of ejected material.
However, this point  will have to be clarified by further studies with a 
sophisticated gravity treatment {\em and} the relevant 
microphysics input.

\item \underline{Weak interactions and neutrino transport}: As has been shown, 
the initial $Y_e$ of the ejected material plays a crucial role in determining
the resulting abundances of the the ejected material. To improve on this aspect
of the problem first of all higher resolution is needed to resolve the low 
density neutron star crust with its higher $Y_e$ values. In addition the weak
interaction physics (including neutrino transport) has to be incorporated 
consistently as well as efficiently to predict
the evolution of $Y_e$ during the coalescence and ejection process.
\end{itemize}
\vspace*{-0.2cm}
Despite these uncertainties it has to be stated that if EOS and gravity 
allow for the ejection of
non-negligible amounts of material and $Y_e$ is close to 0.1, which is a very
reasonable value for neutron star material, neutron star mergers will
contribute r-process nuclei in solar abundances for nuclei with A $>$ 130 and
strongly underabundant amounts of nuclei with  A $<$ 130. 
This is in excellent agreement with recent observations of metal-poor stars
\cite{sneden2000a,sneden2000b} that show a remarkable concordance of heavy
r-process nuclei with scaled solar r-process abundances that breaks down 
for nuclei below $A \approx 130$.\\
Recently, observations of metal-poor stars have been used to argue against 
the possibility of NSMs as substantial sources of Galactic r-process material
\cite{qian2000}.
On average, each volume of the interstellar medium containing $\approx 
3 \times 10^4$M$_\odot$ is enriched by $10^3$ supernovae (SNe) in order
 to obtain solar metallicities. On the other hand, the typical mass of 
the interstellar medium  (ISM) into which the ejecta of an SN remnant are
 mixed (before the remnant shock is quenched) is of the order $3 \times 
 10^4$M$_\odot$ as well. Thus, if the r-process would originate from SNe,
the abundances in the affected ISM resulting from a single SN event in prior
 unpolluted regions would amount to (r/H)$_{SN} \approx 
10^{-3}$(r/H)$_{\odot}$. Since NSM rates are lower than SN rates by 
approximately three orders of magnitude this ratio would in the case of 
a NSM origin be 0 in unpolluted ISM and close to solar in matter which
experienced  NSM ejecta, provided that the mass of the mixed ISM is similar
 to SN remnants. Such behaviour is not observed in low metallicity stars
and therefore NSMs would be ruled out as candidate sites for the r-process.\\
The arguments raised in \cite{qian2000}, however, are too simplistic since 
they omit two important aspects. (1) The amount of NSM ejecta mixing with 
the ISM
is not clear (total energetics and ring-like rather than spherical ejection).
(2) The more important point is that the low-metallicity observations measure
the abundances in individual stars rather than in mixed remnants. 
 Only {\em if} the 
same mass would be swept up and
{\em if} stars would form from fully mixed regions the above argument against 
NSMs holds.
The mixing process of elemental yields from core collapse supernovae with the
 ISM has recently been simulated with a 3D stochastic model  in
order to understand the evolution of the scatter in [element/Fe] ratios during
the history of the Galaxy \cite{argast99}. According to this work the mixing
proceeds via the following stages: for [Fe/H] $< -3.0$ (corresponding to a
Galactic  age of $\sim 200$ Myrs; \cite{argast99} Fig. 6) the ISM is not 
mixed, but rather dominated by local inhomogeneities leading to a very large 
scatter in the observed abundances of individual stars, in agreement with
observations. For [Fe/H] $>$ -2.0 $(\sim 1500$
 Myrs) the ISM is well-mixed (small scatter in observations), inbetween a
continuous transition between both regimes is encountered. 
It could also be shown that {\em a stochastic choice of star formation 
can lead to much smaller metallicities than expected from a well mixed 
remnant}.
For events which occur with a much smaller  frequency (like NSM) this
mixed phase should be delayed to larger metallicities. In addition,
one expects with the large amounts of r-ejecta from NSMs a much larger
scatter than for SNe.
Both effects are seen in the r/Fe observations (a scatter of more than
 a factor of 100 at low metallicities which still amounts to a factor of 10 
at [Fe/H]=-1, see Fig. 2 in \cite{truran2000}).
This fact combined with the suppression of abundances below A=130 in
low-metallicity stars, can be taken as {\em supportive features to a fission 
cycling r-process from a low frequency source such as NSMs}. 
Galactic evolution calculations as in \cite{argast99} are needed in order
 to test the expected amount of scatter as a function of metallicity.

\end{document}